\documentclass[11pt,ocg]{article}
\pdfoutput=1
\usepackage{pgf}
\usepackage{color}
\usepackage{graphicx}
\usepackage{amsmath}
\usepackage{amssymb}
\usepackage{setspace}
\usepackage{fancyhdr}
\usepackage{fancybox}

\usepackage{comment}
\usepackage{tikz,scalefnt,color}
\usepackage{url}
\usetikzlibrary{%
  arrows,%
  shapes.misc,
  shapes.arrows,%
  chains,%
  matrix,%
  positioning,
  scopes,%
  automata}

\newcount\hh
\newcount\mm
\mm=\time
\hh=\time
\divide\hh by 60
\divide\mm by 60
\multiply\mm by 60
\mm=-\mm
\advance\mm by \time
\def\hhmm{\number\hh:\ifnum\mm<10{}0\fi\number\mm}

\newtheorem{theorem}{Theorem}[section]

\newtheorem{proposition}[theorem]{Proposition}

\newtheorem{definition}[theorem]{Definition}

\newtheorem{example}[theorem]{Example}

\def\WTC{\widetilde{C}}
\def\WTA{\widetilde{A}}

\def\Ex{\mathbf{E}}

\def\bU{\mathbb{U}}
\def\bH{\mathbb{H}}

\def\bI{\mathbb{I}}

\def\bM{\mathbb{M}}
\def\bR{\mathbb{R}}

\def\bQ{\mathbb{Q}}
\def\bP{\mathbb{P}}
\def\bK{\mathbb{K}}
\def\bS{\mathbb{S}}
\def\bG{\mathbb{G}}

\def\bA{\mathbb{A}}
\def\bM{\mathbb{M}}
\def\bX{\mathbb{X}}
\def\bY{\mathbb{Y}}
\def\CC{{\mathcal{C}}}
\def\CN{{\mathcal{N}}}
\def\MD{{\mathcal{D}}}
\def\CU{{\mathcal{U}}}

\def\CM{{\mathcal{M}}}
\def\MB{{\mathcal{B}}}
\def\CK{{\mathcal{K}}}
\def\CA{{\mathcal{A}}}
\def\ML{{\mathcal{L}}}
\def\MV{{\mathcal{V}}}
\def\CR{{\mathcal{R}}}
\newcommand{\auto}{\mathsf{T}}

\newcommand\len[1]{{\left\lvert#1\right\rvert}}
\newcommand{\normtt}[1]{\textnormal{\texttt{#1}}}
{\begin{center}\begin{Sbox}\begin{minipage}{\linewidth}\begin{small}\sf\color{red}(#1)}%
{\end{small}\end{minipage}\end{Sbox}\end{center}\colorbox{yellow!90!black}{\TheSbox}}
{\begin{center}\begin{Sbox}\begin{minipage}{\linewidth}\begin{small}\sf\color{black}(#1)}%
{\end{small}\end{minipage}\end{Sbox}\end{center}\colorbox{green!90!black}{\TheSbox}}
{\begin{center}\begin{Sbox}\begin{minipage}{\linewidth}\begin{small}\sf\color{black}(#1)}%
{\end{small}\end{minipage}\end{Sbox}\end{center}\colorbox{orange!90!black}{\TheSbox}}

\DeclareMathOperator{\Pref}{Pref}

\def\alph{\mathcal{A}}

\title{Revisiting Waiting Times in 
DNA Evolution\footnote{\scriptsize This work received support of the ANR project MAGNUM number ANR BLAN- 0204 07}}

\author{Pierre Nicod\`eme\\[1ex]
{\small LIPN - Team CALIN, CNRS-UMR 7030, University Paris North,}\\
{\small Institut Galil\'ee , 99, Avenue Jean-Baptiste Cl\'ement, 93430 -
Villetaneuse, France}\\
{\footnotesize phone: +33 (0)14940-4069,
fax: +33 (0)14826-0712, \ 
{\tt pierre.nicodeme@lipn.univ-paris13.fr}}.}

\begin{document}

\maketitle
\begin{abstract}
Transcription factors are short stretches of DNA (or $k$-mers) mainly
located in promoters sequences that enhance or repress gene expression.
With respect to an initial distribution of letters on the DNA alphabet,
Behrens and Vingron~\cite{BehVin2010} consider a random sequence
of length $n$ that does not contain a given $k$-mer or word of size $k$.
Under an evolution model of the DNA, they compute the probability
$\mathfrak{p}_n$ that this $k$-mer appears after a unit time of 20 years.
They prove that the waiting time for the first apparition of the $k$-mer
is well approximated by $T_n=1/\mathfrak{p}_n$. Their work relies on
the simplifying assumption that the $k$-mer is not self-overlapping.
They observe in particular that the waiting time is mostly
driven by the initial distribution of letters.
 Behrens {\it et al.}~\cite{BehNicNic2012} use an approach by automata
that relaxes the assumption related to words overlaps. Their numerical
evaluations confirms the validity of Behrens and Vingron approach for
non self-overlapping words, but provides  up to 44\% corrections for
highly self-overlapping words such as $\mathtt{AAAAA}$.
We devised an approach of the
problem by clump analysis and generating functions; this approach 
leads to prove
a quasi-linear behaviour of $\mathfrak{p}_n$ for a large range of
values of $n$, an important result for DNA evolution. 
We present here this clump analysis, first by language decomposition, and next by an automaton construction; finally, we describe an
equivalent approach
by construction of Markov automata.
\end{abstract}

\section{Introduction}

Several theoretical studies have tried to give a probabilistic
explanation for the speed of changes in transcriptional gene
regulation (e.g. \cite{stone}, \cite{durrett}).
 Behrens and Vingron~\cite{BehVin2010}
infer how long one has to wait until a given Transcription Factor 
(TF for short) binding site emerges
at random in a promoter sequence. Using a Bernoulli probabilistic model denoted by M0 and estimating evolutionary substitution rates based on multiple species promoter alignments for the three species {\it Homo sapiens}, {\it Pan troglodytes} and {\it Macaca mulatta}, they compute the expected waiting time for every $k$-mer, $k$ ranging from 5 to 10, until it appears in a human promoter. They conclude that the waiting time for a TF binding site is highly determined by its composition and that indeed TF binding sites can appear rapidly, i.e.~in a time span below the speciation time of human and chimp.

However, in their approach, Behrens and Vingron~\cite{BehVin2010} rely on the assumption
that if a $k$-mer of interest appears more than once in a promoter
sequence, it does not overlap with itself. This particularly affects
the waiting times for highly autocorrelated words like e.g. {\tt
  AAAAA} or {\tt CTCTCTCTCT}. Using automata, 
Behrens {\it et al.}~\cite{BehNicNic2012}  relaxed this assumption
and, thus, more accurately compute the expected waiting times until
appearance of $k$-mers
  in a promoter of length 1000.

While Behrens and Vingron~\cite{BehVin2010} and all preceding works
were mostly interested in sequences of fixed length $n=1000$,
   Behrens {\it et al.}~\cite{BehNicNic2012}
  realized that
$\mathfrak{p}_n$ behaves asymptotically linearly with $n$ for a wide
range of lengths. This observation followed from a singularity
analysis performed by the author of the present article;
this property is biologically important, since the lengths
of promoters are in an approximate  range from 1000 base pairs to
2000 base pairs; moreover, it cannot be deduced easily
from the rigorous computations by automata of Behrens {\it et
  al.}~\cite{BehNicNic2012}. We give here  proofs of this property
 that are based on clump analysis 
and use either  combinatorics and language decompositions
or  automata constructions. Our adaptation of previous methods
is new and has theoretical and practical interests.

We present the model in Section~\ref{sec:prelim}. 
We recall in Section~\ref{sec:prevwork} the Behrens-Vingron equations
(2010) and  the automaton approach of Behrens {\it
  et al.} (2012).
The main part of the article is devoted in Section~\ref{sec:clumps}
to  counting the number $H_n$ of putative-hit positions in random
sequences
of length $n$;
at first order, the probability 
$\mathfrak{p}_n$ is then a linear function of $H_n$.
We provide in this section
 the background for the
Guibas-Odlyzko language decomposition and its extension to clump
analysis, and a parallel construction by automata.
  This section also contains
 the translation to generating functions of the formal
languages used and states our result of quasi-linearity of
$\mathfrak{p}_n$; the proof of this result is given in
Section~\ref{sec:lingen} of the Appendix.
Section~\ref{sec:markovauto} sketches a proof by automaton
that does not rely on clump constructions.
\begin{table}[ht!]
A) Estimations for $\nu(a)$, $a\in\mathcal{A}$:\\[0.2cm]
\begin{tabular}{|cccc|}
\hline
$\nu(A)$ & $\nu(C)$ & $\nu(G)$ & $\nu(T)$\\
\hline
0.23889 & 0.26242 & 0.25865 & 0.24004\\
 \hline
\end{tabular}\\

B) Estimations for $p_{\alpha,\beta}(1)$, $\alpha,\beta\in\mathcal{A}$:\\[0.2cm]
\begin{tabular}{|c|cccc|}
\hline
& A & C & G & T\\
\hline
A & 9.99999996e-01& 4.54999995e-09& 1.57499996e-08&3.40000002e-09\\
C & 6.14999993e-09& 9.99999996e-01& 7.14999985e-09&2.17499994e-08\\
G & 2.17499994e-08& 7.14999985e-09& 9.99999996e-01&6.14999993e-09\\
T & 3.40000002e-09& 1.57499996e-08& 4.54999995e-09& 9.99999998e-01\\
\hline
\end{tabular}
\caption{{\bf Parameter estimations.} Numbers taken from \cite{BehVin2010}.\label{estimations}}
\end{table}

\section{Preliminaries}
\label{sec:prelim}
Throughout the article, we assume that promoter sequences evolve according to model M0 which has been described by \cite{BehVin2010}.
\paragraph{Model M0.}
Given an alphabet $\mathcal{A}=\{\mathtt{A,C,G,T}\}$, let
$S(0)=(S_1(0),\dots,$ $ S_n(0))$ denote the initial promoter sequence of
length $n$ taking values in this alphabet. We assume that the letters in $S(0)$ are independent and identically distributed with $\nu(x):=\Pr(S_1(0)=x)$. 
Let the time evolution $(S(t))_{t\geq 0}$ of the promoter sequence be
governed by the $4\times4$ infinitesimal rate matrix
$\bQ=(r_{\alpha,\beta})_{\alpha,\beta\in\mathcal{A}}$. 
The matrix
$\bP(t)=(p_{\alpha,\beta}(t))_{\alpha,\beta\in\mathcal{A}}$ containing
the transitions probabilities of $\alpha$ evolving into $\beta$ in
finite time $t\geq 0$, ($\alpha,\beta\in\mathcal{A}$), can be computed
by $\bP(t)=e^{t\bQ}$; see Karlin-Taylor~\cite{KarTay75}, p. 150-152.
Table~\ref{estimations} provides the parameters used.

\paragraph{The expected waiting time.}
Given a binding site \begin{equation}b=(b_1,\dots,b_k)\quad\text{where } b_1,\dots,b_k\in\mathcal{A},\end{equation} the aim is to determine the expected waiting time until $b$ emerges in a promoter sequence of length $n$ provided that it does not appear in the initial promoter sequence $S(0)$.
More precisely, let
\begin{align*}
T_n=\inf\{t\in\mathbb{N} :&\quad \exists i\in\{1,\dots,n-k+1\}\\
    &\quad\text{ such that }(S_i(t),\dots,S_{i+k-1}(t))=(b_1,\dots,b_k)\}.
\end{align*}
Then, given that $\Pr(b\text{ occurs in }S(0))=0$, the waiting time $T_n$ has
approximately a geometric distribution with parameter $\mathfrak{p}_n$ verifying
\begin{align}
\mathfrak{p}_n&=\Pr(b\text{ occurs in generation 1}\ |\ b\text{ does not
  occur in generation 0})\\
\nonumber &= \Pr(b\in S(1) \ |\ b\not\in S(0)).
\end{align}
See~\cite{BehVin2010}. In particular, one has
${\displaystyle
 \Ex(T_n)\approx\frac{1}{\mathfrak{p}_n}.}$

\section{Previous work}
\label{sec:prevwork}
We present briefly Behrens and Vingron~\cite{BehVin2010} and
Behrens {\it et al.}~\cite{BehNicNic2012} methods.

\subsection{Behrens-$\!$Vingron (2010)}
\label{sec:BehVin}
Considering the $k$-mer $b=b_1\dots b_k$, Behrens and Vingron consider (i)
the
probability that it occurs at time $t=1$ in $S(1)$ and (ii) the
probability
that it
evolves
from a $k$-mer of $S(0)$. Case (i) is computed by inclusion-exclusion
and by assuming that the word $b$ is not self-overlapping.
This gives~\footnote{We use the  notation $|w|_u$ to note the
 number of occurrences of a word $u$ in a word $w$.}
\[Pr(|S(1)|_b \geq 1 )=
\sum_{\ell=1}^{\lfloor n/k
  \rfloor}(-1)^{\ell+1}\binom{n-(k-1)l}{l}\Pr(|S(1)|_b=\ell).\]
Taking in account the evolution probability, they consider next the
words at substitution distance $1$ to $k$ of $b$. Assuming that
the insertion of such words within a sequence $S(0)$ with no occurrence of
$b$ does not create an occurrence of $b$
(this is wrong in general, but a good approximation for non-overlapping words long enough in
a 4 letters alphabet), they obtain
\begin{align}
\label{eq:BehVin}
&\Pr\big(\ |S(1)|_b\geq 1 \ \big|\ |S(0)|_b=0\ \big)\approx
     \sum_{\ell=1}^{\lfloor n/k
  \rfloor}(-1)^{\ell+1}\binom{n-(k-1)l}{l}p_{\ell}, \\
\label{eq:pel}
&\text{with }\quad p_{\ell}=
\left(\sum_{(a_1\dots a_k)\in \CA^k\setminus\{b_1\dots
  b_k\}}\nu(a_1)\dots\nu(a_k)
  \prod_{i=1}^k p_{a_i,b_i}(1)\right)^l.
\end{align}
In the last equation, $p_{\ell}$ is the approximate probability that
$b$
occurs $\ell$ times in $S(1)$ while not occurring in $S(0)$.

\subsection{Automaton approach of Behrens {\it et al.} (2012)}
\label{sec:BeNiNi}
 Behrens {\it et
  al.}~\cite{BehNicNic2012} use the following algorithm.
Let  $\mathsf{A}_b=(Q\!\!:=\!\!\{0,1,\dots,k\},\delta_b,$ $0,\{k\})$
 be the Knuth-Morris-Pratt automaton over the alphabet $\CA$ that recognizes
the
language $\CA^{\star}b\CA^{\star}$. The language $\CA^{\star}\!\setminus\!
\CA^{\star}b\CA^{\star}$
is recognized by the automaton
$\overline{\mathsf{A}}_b=(Q,\delta_b,0,\{0,1,\dots,k-1\})$. 
They construct a product automaton
$\mathsf{P}=\overline{\mathsf{A}}_b\otimes\mathsf{A}_b $ on $\CA\times\CA$ such that\\[-1ex]
\[\mathsf{P}=(Q\times Q,\delta,(0,0),F), \text{ with }
    \left\{\begin{array}{l}
    \delta((p,q),(\alpha,\beta))=(\delta_b(p,\alpha),\delta_b(q,\beta))\\[1ex]
    F=\{0,1,\dots,k-1\}\times \{k\}.
    \end{array}\right.\]
\ \\[-1ex]
They weight (i) any transition $q\xrightarrow{a} q'$ of $\overline{\mathsf{A}}_b$ by $\nu(a)$
and (ii) any transition $c\xrightarrow{(a,a')}c'$ of $\mathsf{P}$ by $\nu(a)\times p_{a\rightarrow a'}$, where $\nu$ is the
initial
distribution of letters and $p_{x\rightarrow y}$ is the probability of
evolution
of letter $x$ to letter $y$ in a unit time. Considering
the corresponding adjacency matrices $\overline{\bA}_b$
and $\bP$,  (provided a suitable
reordering of the lines and columns of the matrices), $V_F$ being a column vector with a one
for each final state in $\mathsf{P}$ and zero elsewhere,
 the probability
$\mathfrak{p}_n$ verifies,\ \\[-3ex]
\[\mathfrak{p}_n=(1,0,\dots,0)\times\bP^n\times V_F\  \Big/ \ 
(1,0,\dots,0)\times\overline{\bA}_b^n\times
(\overbrace{1,\dots,1}^{k\text{ \scriptsize times}},0)^t.\]
\normalsize   
\begin{table}[ht]
\begin{tabular}{|c||rr||rr||c|}\hline
  &  \multicolumn{2}{|c||}{BNN} & \multicolumn{2}{|c||}{BV} &  \\
  &$\Ex_{\operatorname{BNN}}(T_{1000})/10^6$& Rank & $\Ex_{\operatorname{BV}}(T_{1000})/10^6$ & Rank & $\frac{\Ex_{\operatorname{BNN}}(T_{1000})}{\Ex_{\operatorname{BV}}(T_{1000})}$ \\\hline
{\tt CCCCC} &  9.105 &      1021 &  6.304 &        1 & 1.44\\
{\tt GGGGG} &  9.570 &      1022 &  6.666 &      142 & 1.44\\
{\tt TTTTT} & 10.401 &      1023 &  7.457 &      993 & 1.39\\
{\tt AAAAA} & 10.656 &      1024 &  7.654 &     1024 & 1.39\\
{\tt CGCGC} &  7.047 &       699 &  6.446 &       11 & 1.09\\
{\tt TCCCC} &  7.076 &       737 &  6.477 &       17 & 1.09\\
{\tt CCCCT} &  7.076 &       738 &  6.477 &       21 & 1.09\\
{\tt GCGCG} &  7.127 &       787 &  6.518 &       31 & 1.09\\
{\tt CTCTC} &  7.263 &       883 &  6.679 &      148 & 1.09\\
{\tt CACAC} &  7.337 &       945 &  6.750 &      217 & 1.09\\
\hline\end{tabular}
\setlength{\unitlength}{1cm}
\begin{picture}(0,0)
\put(0,-4.6){\rule{\textwidth}{0.2mm}}
\end{picture}
\caption{\label{tab:corrank5} {\bf Expected waiting times
    (generations)
for 5-mers in model M0 with
$\frac{\Ex_{\operatorname{BNN}}(T_{1000})}{\Ex_{\operatorname{BV}}(T_{1000})}>1.09$.}
  (Top 10 results from Table 2 of Behrens {\it et al.}~\cite{BehNicNic2012}).
  $\Ex_{\operatorname{BV}}(T_{1000})$ denotes the expected waiting
  time according to Behrens-Vingron~\cite{BehVin2010} (BV) and
  $\Ex_{\operatorname{BNN}}(T_{1000})$ according to the automaton
  approach of Behrens {\it et al.}~\cite{BehNicNic2012} (BNN). Ranks refer to $5$-mers sorted by their waiting time of appearance according to the two different procedures BV and BNN; rank 1 is assigned to the fastest evolving 5-mer, rank 1024 (=$4^5$) to the slowest emerging 5-mer.}
\end{table}
Table~\ref{tab:corrank5} provides the top 10 $5$-mers with respect
with the correction done by Behrens {\it et al.} (2012) with respect
to Behrens-Vingron (2010).

Considering the minimal period $m(b)$ of a  $k$-mer $b$, such that
\[ m(b) = \min(i,\  |u|=i;  \quad b=u^i.v,\ v \text{ prefix of }u),\]
and noting $i$-periodic a word with minimal period $i$,
 half of the 5-mers, two-thirds of the 7-mers and all of the 10-mers
 with
 $\frac{\Ex_{\operatorname{BNN}}(T_{1000})}{\Ex_{\operatorname{BV}}(T_{1000})}>1.05$
 are either 1- or 2-periodic, i.e. show a high degree of
 autocorrelation.
This implies that, for only $4\%$ of the $5$-mers, $0.2\%$ of the $7$-mers
and $0.002\%$ of the $10$-mers, the exact computations of Behrens {\it
  et al.} (2012) differ by more than $5\%$ of the approximate computations
of Behrens-Vingron (2010). However, as shown in Behrens {\it et al.}
(2012),
a non negligible number of Transcription Factors are highly correlated.

\section{Clump approach}
\label{sec:clumps}
Table~\ref{tab:corrank5} shows clearly the importance of
autocorrelation.
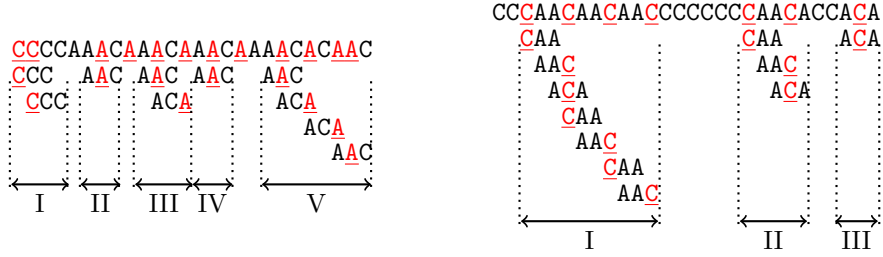
\begin{figure}[!htbp]

\newcommand{\mk}[1] {\textcolor{red}{\underline{#1}}}
\newcommand{\dln}[1] {\draw[dotted,thick](#1,-.4) -- (#1,-1.87);}
\newcommand{\dlnb}[1] {\draw[dotted,thick](#1,-.4) -- (#1,-2.87);}
\ \vskip-5ex
{\scalefont{0.9}
\begin{minipage}{0.5\textwidth}
\begin{center}
\begin{tikzpicture}[scale=0.98]
\draw (0,0) node[anchor=west] (main)       {\texttt{\mk{C}\mk{C}CCAA\mk{A}C\mk{A}A\mk{A}C\mk{A}A\mk{A}C\mk{A}AA\mk{A}C\mk{A}C\mk{A}\mk{A}C}};
\draw (0,-.7) node[anchor=west] (below2)
      {\texttt{\ \mk{C}CC\ \ \ \ \ \ AC\mk{A}\ \ \ \ \ \ AC\mk{A}}};
\draw (0,-1.05) node[anchor=west] (below3)
      {\texttt{\ \ \ \ \ \ \ \ \ \ \ \ \ \ \ \ \ \ \ \ \ AC\mk{A}}};
\draw (0,-1.4) node[anchor=west] (below4)  {\texttt{\ \ \ \ \ \ \ \ \ \ \ \ \ \ \ \ \ \ \ \ \ \ \ A\mk{A}C}};
\draw (0,-.35) node[anchor=west] (below1)   {\texttt{\mk{C}CC\ \ A\mk{A}C\ A\mk{A}C\ A\mk{A}C\ \ A\mk{A}C}};

\dln{0.1}
\dln{0.88}
\draw[<->,thick](0.12,-1.8) -- node[below]{I}  (0.90,-1.8);

\dln{1.05}
\dln{1.58}
\draw[<->,thick](1.06,-1.8) -- node[below]{II}  (1.60,-1.8);

\dln{1.78}
\dln{2.56}
\draw[<->,thick](1.80,-1.8) -- node[below]{III}  (2.58,-1.8);

\dln{3.12}
\draw[<->,thick](2.58,-1.8) -- node[below]{IV}  (3.12,-1.8);

\dln{3.51}
\dln{5.0}
\draw[<->,thick](3.52,-1.8) -- node[below]{V}  (5.0,-1.8);

\end{tikzpicture}
\end{center}
\end{minipage}
}
{\scalefont{0.9}
\begin{minipage}{0.5\textwidth}
\begin{center}
\begin{tikzpicture}[scale=0.98]
\draw (0,0) node[anchor=west] (main)       {\texttt{CC\mk{C}AA\mk{C}AA\mk{C}AA\mk{C}CCCCCC\mk{C}AA\mk{C}ACCA\mk{C}A}};
\draw (0,-.35) node[anchor=west] (below1)   {\texttt{\ \ \mk{C}AA\ \ \ \ \ \ \ \ \ \ \ \ \ \mk{C}AA\ \ \ \ A\mk{C}A}};
\draw (0,-.7) node[anchor=west] (below2)   {\texttt{\ \ \ AA\mk{C}\ \ \ \ \ \ \ \ \ \ \ \ \ AA\mk{C}}};
\draw (0,-1.05) node[anchor=west] (below2)   {\texttt{\ \ \ \ A\mk{C}A\ \ \ \ \ \ \ \ \ \ \ \ \ A\mk{C}A}};
\draw (0,-1.4) node[anchor=west] (below4)  {\texttt{\ \ \ \ \ \mk{C}AA}};
\draw (0,-1.75) node[anchor=west] (below5)  {\texttt{\ \ \ \ \ \ AA\mk{C}}};
\draw (0,-2.1) node[anchor=west] (below6)  {\texttt{\ \ \ \ \ \ \ \ \mk{C}AA}};
\draw (0,-2.45) node[anchor=west] (below7)  {\texttt{\ \ \ \ \ \ \ \ \ AA\mk{C}}};

\dlnb{.5}
\dlnb{2.4}
\draw[<->,thick](.52,-2.8) -- node[below]{I}  (2.38,-2.8);

\dlnb{3.47}
\dlnb{4.42}
\draw[<->,thick](3.49,-2.8) -- node[below]{II}  (4.4,-2.8);

\dlnb{4.8}
\dlnb{5.38}
\draw[<->,thick](4.78,-2.8) -- node[below]{III}  (5.36,-2.8);

\end{tikzpicture}
\end{center}
\end{minipage}
}
\setlength{\unitlength}{1cm}
\begin{picture}(0,0)
\put(0,-6.4){\rule{\textwidth}{0.2mm}}
\end{picture}
\caption{\label{fig:AAA-ACC}{\bf Clumps and putative-hit positions.}
Sequences $S_b(0)$ for $b=\texttt{ACC}$ (left) and
  $S_{b'}(0)$ for $b'=\texttt{AAA}$ (right). The sequence $S_b(0)$ (resp. $S_{b'}(0)$)
  avoids
the $k$-mer $b$ (resp. $b'$). Putative-hit positions are underlined
and in red.
Clumps are shown at their respective position under the
sequences. Note
that extensions to the right of clumps of the set $d(\texttt{AAA})$ for $b'=\texttt{AAA}$, while creating a new
occurrence of a word of the set,
do not add necessarily a new putative-hit position; clump I (right)
contains $7$ occurrences of $d(\texttt{AAA})$, but only $4$ putative-hit
positions for $b'=\texttt{AAA}$. Therefore the number of word
occurrences is not the relevant statistics for precisely counting
putative-hit positions. Note also in the clump I for $b=\texttt{ACC}$ (left) 
that, when the right extension of
a
clump adds a new putative-hit position, this position is not
necessarily in the
extension,
but possibly backwards left.}
\end{figure}
Assuming a four letters alphabet with a uniform probability
distribution,
founding an occurrence of $\mathtt{AAAAA}$ at a position, up to
boundary effects, we have a probability $1/4$ of finding an occurrence
shifted by one position. In contrast, considering an occurrence
of $\mathtt{AACCC}$, we need reading at least $5$ new letters to find
a new occurrence, and the probability of finding two consecutive
occurrences is $1/4^5$. This is a well known fact in combinatorics
of words; words occur by clumps and, while clumps of a non-overlapping
word have only one occurrence of the word, clumps of an overlapping 
word may have several; since the probability (in a uniform model)
of occurrence of any word of a given size at any position 
is the same, the proportion
of text covered by clumps of a non-overlapping word will be larger
than this of an self-overlapping word. This property extends to sets
of words depending of their self-overlap  structure.

We show here that the number of positions in $S(0)$ that can mutate
and provide an occurrence of a $k$-mer $b$ in $S(1)$, or putative-hit
positions, is not a function
of the number of occurrences of $b$ in $S(1)$ or of the neighbours
of $b$ in $S(0)$, but that this number can be computed by a variant
of clump analysis.

\paragraph{Notations.}
Given a word $b$, we note $d(b)$  the set of
its neighbours at edit distance $1$ (by substitution of one letter), 
and $d_{\ell}(b)$ the vector resulting of a  lexicographic  sort of $d(b)$.
Therefore, for an alphabet $\mathcal{A}=\{\texttt{A},\texttt{C}\}$, we have
$d(\texttt{ACC})=\{\texttt{CCC},\texttt{AAC},\texttt{ACA}\}$ and $d_{\ell}(\texttt{ACC})=(\texttt{AAC},\texttt{ACA},\texttt{CCC})$.

\noindent
The  correlation set $\mathcal{C}_{v_1, v_2}$ of two words $v_1$ and
$v_2$ is defined as usual,
\[\mathcal{C}_{v_1, v_2} = \{\  e \ \, | \ \,
\mbox{there exists} \  e' \in \alph^+ \  \mbox{such that} \  
v_1 e =e'v_2 \  \mbox{with} \  \len{e} < \len{v_2}\  \}.
\]  
When we have
$w=v_1=v_2$, we get $\mathcal{C}_{w,w}=\CC$ (the autocorrelation of $w$).

\paragraph{Putative-hit positions.}
Given a sequence $S(0)$ not containing a $k$-mer $b$, a putative-hit
position is any position of $S(0)$ that
can lead by a mutation to an occurrence of $b$ in $S(1)$, where we
assume that a single mutation has occurred.
We have for instance
\[ S(0)=\texttt{CCCAACAC},\quad b=\texttt{ACC}\qquad 
\text{\Large$\leadsto$\normalsize}
  \qquad\underline{S}(0)=\texttt{{\color{red}\underline{C}}CCA{\color{red}\underline{A}}CAC},\]
where the putative-hit positions are underlined in $\underline{S}(0)$.
Mutating any single putative-hit position of $\underline{S}(0)$ leads to a sequence
$S(1)$ with an occurrence of $b=\texttt{ACC}$.

Examples of sequences $S(0)$ for the $3$-mers $\texttt{ACC}$ and $\texttt{AAA}$
 (see Figure~\ref{fig:AAA-ACC}) reveal that
the right method to carry on the computation of putative-hit positions
is clump analysis~\cite{BaClFaNi08}.

\paragraph{Aim of the computation.}
In the following, $H_n$ is the random variable counting the number of
putative-hit positions in a random sequence of length $n$.
We consider the generating function $F_b(z,t)$ that counts the
number of putative-hit
 positions for the $k$-mer $b$ in texts avoiding this $k$-mer,
\begin{equation}
\label{eq:gfputpos}
F_b(z,t)=\sum_{w\in\overline{\mathcal{A}^{\star}_b}}\Pr(w)z^{|w|}
t^{\operatorname{put-hit-pos}(w)},
\end{equation}
where $\overline{\mathcal{A}^{\star}_b}$ is the set of sequences of any
length 
that do not contain the $k$-mer $b$ and
$\operatorname{put-hit-pos}(w)$ is the number of putative-hit
positions of the word $w$. Note that, up to probability
of second order small magnitude,
only one putative-hit position will mutate.

\subsection{Analysis ``\`a la Guibas-Odlyzko''}
\label{sec:RS}
Considering a reduced set of words (no word is factor
of another word in the set),  R\'egnier and Szpankowski~\cite{ReSz97,ReSz98b}  and
R\'egnier~\cite{Regnier00b}
use (as an evolution of Guibas and Odlyzko
previous work~\cite{GuiOdl81a,Guiodl81b}) a
natural parsing or decomposition of texts with respect to the
occurrences
of the set.

We follow here the corresponding presentation of Lothaire~\cite{Lot05} (Chapter 7).
Let $\mathcal{V}=\{v_1,\dots,v_r\}$ be a reduced set of words. We have,
formally
\small
\begin{definition}{Right, Minimal, Ultimate and Not languages.}
\label{def:RMUlang}
\begin{itemize}\addtolength{\itemsep}{-0.50\baselineskip}
\addtolength{\topsep}{-1.90\baselineskip}
\item[--] The ``Right'' language $\mathcal{R}_i$ associated to the
  word $v_i$ is the set of words
$ 
\mathcal{R}_i=\{r\, |\, r=e\cdot v_i \mbox{ and there is no } \upsilon\in
  \mathcal{V} \mbox{ such that } r=x\upsilon y \mbox{ with } \len{y}>0\}.
$
\item[--] The ``Minimal'' language $\mathcal{M}_{ij}$ leading from a
  word $v_i$ to a word $v_j$ is the set of words
 $
  \mathcal{M}_{ij}=\{m\, | \, v_i \cdot m=e \cdot v_j \mbox{ and there
    is no } \upsilon\in \mathcal{V} \mbox{ such that } v_i \cdot
  m=x\upsilon y \mbox{ with } \len{x}>0, \len{y}>0\}.
$
\addtolength{\topsep}{-0.90\baselineskip}
\item[--] The ``Ultimate'' language of words following the last
  occurrence of the word $v_i$ (such that this occurrence is the last
occurrence of $\MV$ in the text) is the set of words
$ 
\mathcal{U}_{i}=\{u\, |\, \mbox{ there is no } \upsilon\in \mathcal{V}
\mbox{ such that } v_i\cdot u=x\upsilon y \mbox{ with } \len{x}>0\}.$
\addtolength{\topsep}{-0.90\baselineskip}
\item[--] The ``Not'' language $\mathcal{N}$ is the set of words with no
occurrences of  $\mathcal{V}$,
$ 
\mathcal{N}=\{n\ |\ \mbox{there is no } \upsilon\in \mathcal{V} \mbox{
  such that } n=x\upsilon y\}.
$
\end{itemize}
\end{definition}
\normalsize
It is possible to obtain the generating functions of these languages
by combinatorics and by new automata constructions.
\subsection{R\'egnier-Szpankowski equations}
\label{sec:regszpeq}
Considering the matrix $\bM=\left(\CM_{ij}\right)$ and
using $\CC_{ij}=\CC_{v_i,v_j}$ as a shorthand, we have
\begin{eqnarray}
\label{eq:lotvw1}
&\displaystyle\bigcup_{k\geq 1}\left(\bM^k\right)_{i,j}=\CA^{\star}\!\cdot\! v_j +
\CC_{ij}-\delta_{ij}\epsilon,
\quad &\displaystyle\CU_i\!\cdot\!\CA=\bigcup_{j} \CM_{ij}+\CU_i-\epsilon, \\
\label{eq:lotvw5}
&\displaystyle\CA\!\cdot\! \CR_j -\left(\CR_j-v_j\right)=\bigcup_{i} v_i\CM_{ij},
\quad &\!\!\!\!\displaystyle\CN\!\cdot\! v_j= \CR_j+\bigcup_{i}
\CR_i\left(\CC_{ij}-\delta_{ij}\epsilon\right),
\end{eqnarray}
\normalsize
where the Kronecker symbol $\delta_{ij}$  is $1$ if $i=j$ and $0$
elsewhere.
These equations are non-ambiguous and translate to generating functions, where for a language
$\ML$ and its generating function $L(z)$, we have
${\displaystyle L(z)=\sum_{w\in \ML} \Pr(w)z^{|w|}.}$
Translating the system of Equations
(\ref{eq:lotvw1},$\,$\ref{eq:lotvw5})
to generating functions and solving the resulting system
provide the generating functions $R_i(z)$, $M_{i,j}(z)$, $U_j(z)$ and
$N(z)$ of the Right, Minimal, Ultimate and Not languages.
The parsing by languages is now reflected in the following equation
\small
\begin{equation}
\label{eq:parseRS}
\frac{1}{1-z}=N(z)+ (R_1(z),\dots,R_r(z))(\bI-\bM(z))^{-1}
         \left(\begin{array}{c}
           U_1(z)\\\vdots\\U_r(z)\end{array}\right)
\end{equation}
where \small $\dfrac{1}{1-z}$ \normalsize is the generating function
of $\CA^{\star}$, the set of all texts.
\subsection{New automata constructions}
The languages $\CR_i,\CM_{ij},\CU_j,\CN$ are recognized by the
following automata (where $\bigotimes$ is the usual automaton
product):
\def\Onot{\operatorname{Not}}
\begin{align*}
&\CR_i= 
  \CA^{\star}.v_i\!\!\!\bigotimes\displaylimits_{s\in\{1,\dots,r\}}\!\!\!\!\Onot(\CA^{\star}v_s\CA)
&&
v_i\CM_{ij}=
 v_i\CA^{\star}\bigotimes\CA^{\star}.v_j\!\!\!\bigotimes\displaylimits_{s\in\{1,\dots,r\}}\!\!\!\!\Onot(\CA^{+}v_s\CA^{+})\\
&v_j\CU_j=
v_j\CA^{\star}\!\!\!\bigotimes\displaylimits_{s\in\{1,\dots,r\}}\!\!\!\!\Onot(\CA^{+}v_s\CA^{\star})
&&\quad\CN=\Onot\left(\bigotimes\displaylimits_{s\in\{1,\dots,r\}}\CA^{\star}v_s\CA^{\star}\right)
\end{align*}

\subsection{Constrained languages}
\label{sec:conslang}
\paragraph{Language approach.}
Considering the word $b=\texttt{AAA}$, and $d_{\ell}(\texttt{AAA})=(\texttt{AAC},\texttt{ACA},\texttt{CAA})$,
 we can compute from the vector of words
$(\texttt{AAC},\texttt{ACA},\texttt{CAA},\texttt{AAA})$ a row vector of Right languages
$(\CR_1,\CR_2,\CR_3,\CR_4)$,
a matrix of Minimal languages $(\CM_{ij})$ with $i$ and $j$ from $1$
to $4$ and a column vector of Ultimate languages
$(\CU_1,\CU_2,\CU_3,\CU_4)^{\mathbf{t}}$.
Extracting the languages with indices from $1$ to $3$ provides us for $d_{\ell}(\texttt{AAA})$
with the 
Right $\overline{\CR}_i=\CR_i$, Minimal $\overline{\CM}_{ij}=\CM_{ij}$
and Ultimate $\overline{\CU}_j=\CU_j$ languages avoiding the word
$\texttt{AAA}$.

The construction given here is fully general. For any
finite
alphabet $\mathcal{A}$ and any word $b$, it is (at least
theoretically)
possible to
solve the R\'egnier-Szpankowski equations for the extended sequence
$(b_1,\dots,b_r,b)$ where $d_{\ell}(b)=(b_1,\dots,b_r)$, which
provides the constrained languages for $d_{\ell}(b)$.
\paragraph{Automata approach.} It is also immediate to construct
by automata the constrained languages. For instance, we have, for
$i,j\in\{1,2,3\}$,
\[v_i\overline{\CM_{ij}}=
 v_i\CA^{\star}\bigotimes\CA^{\star}.v_j\!\!\!\bigotimes\displaylimits_{s\in\{1,2,3\}}\!\!\!\!\Onot(\CA^{+}v_s\CA^{+})\bigotimes
 \Onot(\CA^{\star}b\CA^{\star}),\]
and the general case follows also easily.

\subsection{Clump equations by language decomposition}
\label{sec:cluback}
Bassino {\it et al.}~\cite{BaClFaNi08} modify the
R\'egnier-Szpankowski analysis of reduced sets to more specifically
consider
clumps of occurrences, where a clump is constituted either ({\it i}$\,$) of a
single isolated (with no overlap with other occurrences)
occurrence of a word of the pattern, or ({\it ii}$\,$) of a maximal set of
occurrences
where each occurrence overlaps at least another one. 

We consider the {\sl residual} language $\MD=\ML.w^{-}$ as $\MD=\{v, \
v\cdot w\in \ML\}$.  Considering two languages $\ML_1$ and $\ML_2$, 
we write $\ML_2-\ML_1=\ML_2\setminus\ML_1 = \{v; \ v\in\ML_2, v\not\in\ML_1\}$.

The clumps can be generated by a matrix of codes $\bK=\left(\CK_{ij}\right)$.
With
\begin{equation}
\label{eq:basiccode}
\CK_{ij}= \MB_{ij}-\MB_{ij}\CA^{+}\quad\text{and}\quad
    \left\{\begin{array}{ll} &\MB_{ij}=\CC_{ij}\cap \CM_{ij} \quad \text{if}\ i\neq j,\\[0.5ex]
                         
                         &\MB_{ii}=(\CC_{ii}-\epsilon)\cap \CM_{ii},\end{array}\right.
\end{equation}
the language decomposition by clumps for a pattern $\mathcal{V}=\{v_1,\dots,v_r\}$ is
\small
\begin{equation}
\label{eq:clump}
\CA^{\star}=\CN+(\CR_1 v_1^{-},\dots,\CR_r v_r^{-})\  \bG \Big((\bM-\bK)^{-}\bG\Big)^{\star}
\begin{pmatrix} \CU_1\\\vdots\\\CU_r \end{pmatrix},\ \text{with}\ 
    \left\{\begin{array}{l}
    \bK=(\CK_{ij}),\ \ \bS=\bK^{\star},\\[1ex]
 \bG=
\begin{pmatrix}
 v_i \bS_{ij}
\end{pmatrix}\end{array}\right.
\end{equation}
\normalsize

\begin{example}
For the word $w=\normtt{AAAA}$, we have
$ \mathcal{C}=\{\epsilon,\normtt{A},\normtt{A},\normtt{AAA}\} \text{  and  }  \mathcal{K}=\{\normtt{A}\}$.
For the pattern $\mathcal{V}=\{\normtt{TATAT},\normtt{CATAT}\}$, we have
$ \mathcal{C}_{\normtt{CATAT,TATAT}}=\{\normtt{AT},$ $\normtt{ATAT}\}$  and  
$\mathcal{K}_{\normtt{CATAT,TATAT}}=\{\normtt{AT}\}$. For the pattern 
$\mathcal{V}'=\{\normtt{CAA},\normtt{AAT},\normtt{AAA}\}$, we have
$\mathcal{C}_{\normtt{CAA},\normtt{AAT}}= \{\normtt{T},\normtt{AT}\}$ and
$\mathcal{K}_{\normtt{CAA},\normtt{AAT}}= \{\normtt{T}\}$.
\end{example}
\paragraph{Constrained clumps.}
The finite code languages generating the correlation languages of two words
are easy to compute directly; one must however also avoid the
forbidden
word $b$ while extending clumps. We therefore define for $v_i$
(resp. $v_j$) the
$i$-th (resp. $j$-th)
entry of the sequence $d_{\ell}(b)$\\[-2ex]
\[ \overline{\CK}_{ij}=\{h\in \CK_{ij};\qquad |v_i.h|_{b}=0\},\]\\[-3ex]
where $|g|_b$ is again the number of occurrences of the word $b$ in the word
$g$. Since the  sets $\CK_{ij}$ are finite, the computations of the
codes
$\overline{\CK}_{ij}$
 can be done by 
string-matching. 

Gathering everything, we obtain a constrained version of Equation
(\ref{eq:clump})
for the language  $\overline{\CA^{\star}_b}$ of texts avoiding the
word $b$,
\\[-2ex]
\begin{equation}
\label{eq:consclump}
\overline{\CA^{\star}_b}=\overline{\CN}+(\overline{\CR}_1 v_1^{-},\dots,\overline{\CR}_r v_r^{-})\  \overline{\bG} \Big((\overline{\bM}-\overline{\bK})^{-}\overline{\bG}\Big)^{\star}\!
\begin{pmatrix} \overline{\CU}_1\\[-1ex]\vdots\\\overline{\CU}_r \end{pmatrix},\ \text{with}\ 
    \left\{\begin{array}{l}
    \overline{\bK}=(\overline{\CK}_{ij}),\\ \overline{\bS}=\overline{\bK}^{\star},\\
 \overline{\bG}=
\begin{pmatrix}
 v_i\overline{\bS}_{ij}
\end{pmatrix}\end{array}\right.
\end{equation}

\subsection{Computing the generating function of the number of putative-hit positions}
\label{sec:gf}
We prove that the computation of the generating function
$F_b(z,t)$ of Equation~\ref{eq:gfputpos} follows from Equation
\eqref{eq:consclump}.
Indeed, taking in consideration the lengths of the words and the number
of occurrences of putative-hit positions, we have first $v_i(z,t)=\Pr(v_i)tz^{|v_i|}$
for each $v_i\in d(b)$.
Next, for each $\overline{\mathcal{K}}_{ij}$, we can compute by string matching the
number of putative-hit positions in each word  of
$v_i.\overline{\mathcal{K}}_{ij}$. This gives
\[ \overline{\mathcal{K}}_{ij}(z,t)=\sum_{w\in \overline{\mathcal{K}}_{ij}}\Pr(w)
      t^{\operatorname{put-hit-pos}(v_i.w)-1}z^{|w|},\]
where we substracted the putative-hit position occurring within $v_i$.

From the last equation and Equation \eqref{eq:consclump}, we get
\begin{align}
\nonumber
 \overline{\bK}(z,t)&=\left(\overline{\mathcal{K}}_{ij}(z,t)\right),
                       \quad
                       \overline{\bS}(z,t)=\left(\bI-\overline{\bK}(z,t)\right)^{-1},\\
\label{eq:KGS}
                        \overline{\bG}(z,t)&=\Big(v_i(z,t)
                       \overline{\bS}_{ij}(z,t)\Big).
\end{align}
Substituting  in
Equation \eqref{eq:consclump} $\overline{\bG}$ by $\overline{\bG}(z,t)$ and 
$\overline{\CN}, \overline{\CR}_i v_i^{-},
(\overline{\bM}\!-\!\overline{\bK})$ and $\overline{\CU}_i$ with $1\!\leq\!
i\!\leq\! r$ by their
translations to generating functions (that depend only of the variable
$z$) provides 
the expression of  $F_b(z,t)$ that has been formally defined in Equation\eqref{eq:gfputpos}.

We also have\ \\[-2ex]
\begin{equation}
\label{eq:fbarn}
F_b(z,1)=\overline{\mathcal{A}^{\star}_b}(z,1)=\sum_{n\geq
  0}\overline{f}_n^{(b)} z^n=\sum_{n\geq 0}\Pr(S_n(0)\not\in
\mathcal{A}^{\star}b\mathcal{A}^{\star})\times z^n,
\end{equation}\ \\[-2ex]
where $\overline{f}_n^{(b)}$ is the probability that a random sequence of length
$n$ does not contain the word $b$.
This implies that the conditionned\footnote{We use the classical equation $\Pr(A|B)=\Pr(A)/\Pr(B)$
 for two events
$A$ and $B$ such that $B\subset A$.} expectation $\Ex(\widetilde{H}_n)$ of the number of putative-hit positions verifies
\begin{equation}
\label{eq:EHN}
\Ex(\widetilde{H}_n)=\Ex(H_n)\Big/ \overline{f}_n^{(b)}=[z^n]\left.\frac{\partial F_b(z,t)}{\partial t}\right|_{t=1} \Big/ \overline{f}_n^{(b)}.
\end{equation}
Considering again the evolution matrix
$\bP(1)=\left(p_{\alpha\rightarrow\beta}\right)$ with
$\alpha,\beta\in\{\mathtt{A,C,G,T}\}$, we state the following
  proposition that we prove in the Appendix, Section~\ref{sec:lingen}.
\begin{proposition}
\label{prop:linear}
For (i) $\max_{\alpha,\beta\in
  \CA; \alpha\neq\beta}\left(p_{\alpha\rightarrow  \beta}\right)\ll 1$ and (ii) $n$ large enough with 
$n\ll \min^{-1}_{\alpha,\beta\in \CA; \alpha\neq\beta}\left(p_{\alpha\rightarrow  \beta}\right)$,
 the probability that a $k$-mer occurs at time $1$ while not occuring
at time $0$ in a sequence of length $n$ behaves quasi-linearly with respect
to the length $n$. The convergence to this quasi-linear
regime is exponential.
\end{proposition}

\begin{figure}[ht!]
\begin{center}
\begin{tikzpicture}[shorten >= 0.1pt, node distance=1.8cm, thick,
    scale=.5]
\node (chi) at (3,3) {$(\chi=16)$};
\node[state,minimum size=3pt,initial]   (q00) at (4,8)   { $0$};
\node[state,minimum size=3pt]           (q01) at (7,9)   { $1$};
\node[state,minimum size=3pt]           (q02) at (10,10) { $2$};
\node[state,minimum size=3pt,accepting] (q03) at (13,11) { $3$};
\node[state,minimum size=3pt]           (q04) at (16,12) { $4$};
\node[state,minimum size=3pt,accepting] (q05) at (19,12) { $5$};
\node[state,minimum size=3pt]           (q06) at (10,7.5){ $6$};
\node[state,minimum size=3pt,accepting] (q07) at (13,7.5){ $7$};
\node[state,minimum size=3pt]           (q08) at (16,9)  { $8$};
\node[state,minimum size=3pt,accepting] (q09) at (19,9)  { $9$};
\node[state,minimum size=3pt]           (qchi) at (7,6.5) {$\chi$};
\node[state,minimum size=3pt]           (q10) at (16,6)  {$10$};
\node[state,minimum size=3pt,accepting] (q11) at (19,6)  {$11$};
\node[state,minimum size=3pt]           (q12) at (10,5)  {$12$};
\node[state,minimum size=3pt,accepting] (q13) at (13,4)  {$13$};
\node[state,minimum size=3pt,accepting] (q14) at (16,3)  {$14$};
\node[state,minimum size=3pt,accepting]           (q15) at (19,3)  {$15$};

\path[->] (q00) edge node[above]{$\mathtt{A}$} (q01)
          (q01) edge node[above]{$\mathtt{A}$} (q02)
          (q02) edge node[above]{$\mathtt{\WTC}$} (q03)
          (q03) edge node[above]{$\mathtt{A}$} (q04)
          (q04) edge node[above]{$\mathtt{A}$} (q05)
          (q04) edge[bend left] node[left, pos=0.2]{$\mathtt{C}$} (q10)
          (q01) edge node[above]{$\mathtt{C}$} (q06)
          (q06) edge node[above]{$\mathtt{\WTA}$} (q07)
          (q07) edge node[above]{$\mathtt{A}$} (q08)
          (q07) edge node[above]{$\mathtt{C}$} (q10)
          (q08) edge node[above]{$\mathtt{\WTC}$} (q09)
          (q09) edge[bend right] node[right, pos=0.2]{$\mathtt{A}$} (q15)
          (q10) edge node[below]{$\mathtt{\WTA}$} (q11)
          (q11) edge[bend right] node[above]{$\mathtt{C}$} (q10)
          (q00) edge node[below]{$\mathtt{C}$} (qchi)
          (qchi) edge[loop below] node[below]{$\mathtt{C}$} (qchi)
          (qchi) edge node[below]{$\mathtt{A}$} (q12)
          (q12) edge node[below]{$\mathtt{\WTA}$} (q13)
          (q12) edge node[left]{$\mathtt{C}$} (q06)
          (q13) edge node[below]{$\mathtt{\WTC}$} (q14)
          (q14) edge node[below]{$\mathtt{A}$} (q15)
          (q15) edge[bend right] node[right]{$\mathtt{A}$} (q05)
          (q15) edge node[below]{$\mathtt{C}$} (q10)
;
\end{tikzpicture}
\end{center}
\setlength{\unitlength}{1cm}
\begin{picture}(0,0)
\put(0,-3.4){\rule{\textwidth}{0.2mm}}
\end{picture}
\caption{\label{fig:autoDaaa} Automaton for constrained clumps of
$d(\mathtt{AAA})=\{\mathtt{AAC,ACA,CAA}\}$. Double circles signals
an occurrence of one of these words. Transitions covered by tildes
($\mathtt{\WTA},\mathtt{\WTC}$) emits a signal counting a putative-hit position.
The missing transitions $\mathtt{A}$ have been erased since we want
to avoid occurrences of $b=\mathtt{AAA}$. The missing transitions
$\mathtt{C}$ point to the state $\chi$. All states are terminal.}
\end{figure}
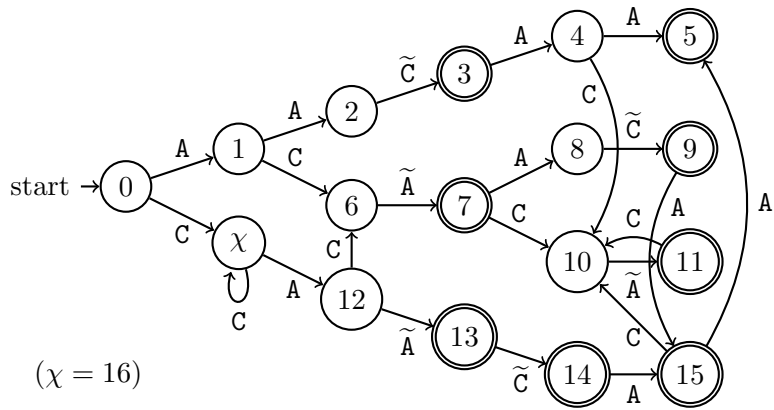

\subsection{Approach by automata of clumps}
\label{sec:clumpauto}
We can alternatively use the construction of clumps by automata given
in Bassino {\it et al.}~\cite{BaClFaNi08}.

For a set $\mathcal{V}=\{v_1, \dots, v_r\}$ with correlation sets
$\CC_{ij}$ we construct a kind of ``Aho-Corasick''
automaton on the following set of words $X$ 
\[
X= \{ v_i \cdot w \ | \ \text{$1\le i \le r$ and $w \in \{\epsilon\} \cup \CC_{ij}$ for some $j$}\}.
\]
The considered automaton $\auto$ is built on
$X$ with set of states $Q=\Pref(X)$  and start or initial state
$s=\epsilon$. The transition function is defined (as in the Aho-Corasick
construction) by 
\[
\delta(p, x) = \quad\text{the longest suffix of $px \in
  \Pref(X)$}.
\] 
We build with this construction, for any $k$-mer $b$, the automaton
recognizing the clumps of the neighbours $d(b)$ of $b$ while
avoiding occurrences of $b$; this last condition can be made effective
by doing an automaton product.
Assuming that the set of states of the resulting automaton $\auto$ 
is $ Q=\{0,1,\dots,s\}$ and that the
initial state is labelled $0$, we set all the states of the automaton
$\auto$ to terminal to recognize all sequences avoiding $b$.
Therefore, we have
\[\auto=(\{0,1,\dots,s\},\delta,0,\{0,1,\dots,s\}.\]
See Figure~\ref{fig:autoDaaa} for an example with the alphabet
$\CA=\{\mathtt{A,C}\}$, the $k$-mer $b=\mathtt{AAA}$
and $d(b)=\{\mathtt{AAC,ACA,CAA}\}$.
Transitions with a ``tilde'' correspond to finding a new putative-hit
position
in the last recognized occurrence of a word of $d(b)$.
\paragraph{Clump-Core.}
We consider the set of states $O$ that recognize an occurrence of $d(b)$,
\[  O=\{ q,\quad \delta(0,w)=q,\ w\in X\}.\]
We also consider the set of states $\overline{E }$ that do not belong
to a clump extension,
\[ \overline{E }= \{q, \quad \delta(0,w)=q, \ w\in \widehat{\Pref}(d(b))\},\]
where $\widehat{\Pref}(d(b))$ is the set of strict prefixes of words of $d(b)$.

We define finally the $\operatorname{Clump-Core}$ of the automaton by
its set of states $E $ which verifies
\[ E =  Q \setminus \overline{E }.\]
Referring to the automaton of Figure~\ref{fig:autoDaaa}, we have
$\overline{E}=\{0,1,2,16\, (\chi),6,12\}$ and $E=\{3,4,5,7,8,9,10,11,13,14,15\}$.
\paragraph{Markov property.}
By construction of the automaton, for any word $w$ with $|w|\leq|b|$,
we have the following property,
\[\forall e\in E,\  \forall w\ \text{with }  (|w|\leq|b|)
\left\{\begin{array}{l}
     \not\exists w'\neq w \text{ with }(|w'|=|w|)\\[1ex]
     \text{such that }\delta(q_1,w)=\delta(q_2,w')=e.
  \end{array}\right.\]
This property can be proved iteratively with respect to the length
of the words.

\paragraph{Handling the putative-hit positions.}
For simplicity, we assume that there is only one type of  mutation,
but the method extends to the  general case. We count as previously
the putative-hit positions by the variable $t$.

For each state $o\in O$ (recognizing an occurrence of $d(b)$),
let $\theta(o)$ be the word $w$ with $|w|\leq|b|$, of maximal length,
and verifying,
\begin{enumerate}
\item there exists $q$ such that $\delta(q,w)=o$,
\item there is no $u\in\widehat{\Pref}(w)$ such that $\delta(q,u)\in  O$.
\end{enumerate}
By the Markovian property, this defines a unique word. Referring to
Figure~\ref{fig:autoDaaa}, we have $\theta(7)=\mathtt{ACA}$,
$\theta(5)=\mathtt{AA}$, $\theta(14)=\mathtt{C}$, and $\theta(15)=\mathtt{A}$.
Moreover, the Markovian property asserts that reading backward $|b|$
transitions from any state $o\in O$ does a reverse spelling of a unique
word of $d(b)$. We can next locate the putative-hit position within
this
word and check if it belongs to $\theta(o)$.

The adjacency matrix $\bH(t)=(h_{ij}(t))$ associated to the automaton $\auto$ is
then defined as follows:
$h_{ij}(t)=0$ if there is no transition from $i$ to $j$; elsewhere,
assume that $\delta(i,\alpha)=j$. We have then
\[
 h_{ij}(t)=\left\{\begin{array}{l}
     \Pr(\alpha) \text{ if }\left| \begin{array}{l}j\not\in  O,\\
       j\in  O \text{ and } \theta(j) 
        \text{ contains no putative-hit position,}
        \end{array}\right.\\[2ex]
     \Pr(\alpha)\times t \text{ elsewhere}.
   \end{array}\right.
\]
The generating function $F_b(z,t)$ defined in Equation
(\ref{eq:gfputpos}) verifies
\begin{align*}
F_b(z,t)&=(1,0,\dots,0)\times\big(\bI+z\bH(t)+\dots+z^n\bH^n(t)
+\dots\big)
\times \mathbf{1}^t\\
&=(1,0,\dots,0)\times(\bI-z\bH(t))^{-1}\times
\mathbf{1}^t.
\end{align*}

\section{Yet another proof by automata}
\label{sec:markovauto}
We sketch a proof that does not make use of clumps. The construction
is computationally very costly.

We build the (pruned) Knuth-Morris-Pratt automaton $\mathsf{K}$ recognizing 
$\overline{\CA^{\star}b\CA^{\star}}$ (the set of sequences
avoiding
the $k$-mer $b$).

Next we compute the order-$(2|b|-1)$ Markov automaton $\mathsf{M}$ 
of $\mathsf{K}$. The transitions of this automaton are words of size
$2|b|$. It is possible by reading the transitions to know when
a new putative-hit position is present, and to multiply the
corresponding
entry in the associated adjacency matrix by the counting variable $t$.
Let $\bM(t)$ be this matrix.
The matrix associated to the automaton $\mathsf{K}$ is positive,
irreducible
and transitive; so is the  matrix $\bM(t)$, disregarding a trie-like
structure leading to its recurrent part. 
\def\pmut{p_{\operatorname{mut}}}
Writing $\pmut$ the probability of mutation, we can make the
substitution $t\leadsto (1-\pmut)+x\times\pmut$.
We then have for the recurrent part $\bR(t)$ of $\bM(t)$,
\[\bU(x) := \bR((1-\pmut)+x\times\pmut)=\bY+x\pmut\bX.\]
Assuming that $n\times\pmut=o(1)$, we get for a polynomial $P(x)$
\begin{equation}
\label{eq:bRn}
\bU^n(x)=\bY^n+xn\pmut\bY^{n-1}\bX+x^2P(x)\times O\left((n\pmut)^2\right).
\end{equation}
Writing $\xi_u$ and $\xi_y$ the dominant eigenvalues of $\bU(1)$ and $\bY$,
the property $n\pmut=o(1)$ entails that
$\xi_r^n=\xi_y^n\times(1+o(1))$.
We then deduce from Equation (\ref{eq:bRn}) that
\[\mathfrak{p}_n\approx\frac{[x^1](1,0,\dots,0)\bU^n(x)\mathbf{1}^t}{(1,0,\dots,0)\bU^n(1)\mathbf{1}^t}
=(\alpha+\beta\!\times\! n) \times(1+o(1)).\] 

\section{Conclusion}
We provided several methods for analysing waiting times
in DNA evolution that give insights in the structure
of the problem. We showed that clump analysis and generating functions
are powerful and
 convenient
tools for this aim and used
either  language analysis methods or automata constructions.
In particular we proved the property of quasi-linearity
related to the probability of first occurrence of a $k$-mer after
a unit time.

\paragraph{Acknowledgements.} We thank Sarah Behrens and Cyril Nicaud
for helpful discussions and technical help.

\bibliographystyle{acm} 
\small  

\appendix

\normalsize
\section{Singularity analysis}
\label{sec:lingen}
The methods developed in Section~\ref{sec:gf} apply to any $k$-mer with
any finite alphabet.
Moreover,  using the additivity of
expectations,
 we could split the putative-hit
positions
along their types; with the toy alphabet $\{\mathtt{A},\mathtt{C}\}$, we would get
two putative-hit positions type, $(\mathtt{A}\rightarrow \mathtt{C})$ and $(\mathtt{C}\rightarrow
\mathtt{A})$.
By following the same footsteps as in Section~\ref{sec:gf}, we can
now compute the expectations of putative-hit positions 
$\Ex\left(H^{(\mathtt{A}\rightarrow \mathtt{C})}_n\right)$ and $\Ex\left(H^{(\mathtt{C}\rightarrow \mathtt{A})}_n\right)$ 
which correspond to the two types of mutation. 
Considering only the case $(\mathtt{A}\rightarrow \mathtt{C})$, we can by
pattern-matching compute 
$\overline{\mathcal{K}}_{ij}\left(z,t_{(\mathtt{A}\rightarrow
  \mathtt{C})}\right).$
We have as previously 
$v_i\left(z,t_{(\mathtt{A}\rightarrow \mathtt{C})}\right)=\Pr(v_i)z^{|v_i|}t_{(\mathtt{A}\rightarrow \mathtt{C})}$.

We write in the following for sake of simplicity $x=t_{(\mathtt{A}\rightarrow \mathtt{C})}$, 
and consider the generating function
$F_b(z,x)=F_b(z,t_{(\mathtt{A}\rightarrow \mathtt{C})})$
where the function $F_b(z,t)$ is defined in
Equation \eqref{eq:gfputpos}.

The solutions of the R\'egnier-Szpankowski equations provide functions
that
are rational~\footnote{This property follows also from an equivalent
  approach
by finite automata and use of the Chomski-Sch\"utzenberger 
algorithm~\cite{NiSaFl02}
that
leads to solve a linear system of equations.}. Similarly, each term of
the matrix  Equation
(\ref{eq:consclump}) is rational and so are the corresponding
extensions to counts of
putative-hit
positions that lead to the explicit value of $F_b(z,x)$.

We can therefore write for two polynomials $P(z,x)$ and $Q(z,x)$
\[ F_b(z,x) = \frac{P(z,x)}{Q(z,x)}\quad\text{and}\quad
F_b(z,1)=\sum_{n\geq 0}\overline{f}_n^{(b)} z^n=\frac{P(z,1)}{Q(z,1)},
\]
where, again,  $\overline{f}_n^{(b)}$ is the probability that a random sequence of
length $n$ has no occurrence of $b$. We have
\[E(z)=\sum_{n\geq 0}\Ex\left(H_n^{(\mathtt{A}\rightarrow\mathtt{C})}\right)z^n
=\frac{P'_x(z,1)}{Q(z,1)}-\frac{P(z,1)Q'_x(z,1)}{Q^2(z,1)}.\]
This series has only positive coefficients and by Pringsheim 
Theorem~\cite{FlajoletSedgewick2009}[Th. IV.6, p.240],
it has a real positive  singularity
on the circle of convergence that we note $\tau$;
by considering the
  automaton
recognizing $\overline{\CA^{\star}b\CA^{\star}}$, the associated
irreducible and primitive matrix, and Perron-Frobenius
properties of positive matrices~\cite{KarTay75}, this real positive singularity is
dominant.
The singularity $\tau$ is also
the smallest positive solution of
$Q(z,1)=0$.

Therefore,
 extracting the $n$th Taylor coefficient of the generating
functions $E(z)$ and $F_b(z,1)$ by
Cauchy integrals along a
circle of radius $\tau<R<\tau_2$, where $\tau_2$ is the value of the
second largest singularity(ies) in modulus, we obtain  for constants $\psi$, $\phi_1$ and $\phi_2$
\[ \overline{f}_n^{(b)} =\psi\times
\tau^{-(n-1)}\left(1+\mathcal{O}\left(B^n\right)\right),\qquad (B<1),\qquad\phantom{1}\]
\begin{equation}
\label{eq:Mn}
\text{and }\qquad \Ex(H_n^{(\mathtt{A}\rightarrow\mathtt{C})}) = [z^n]E(z)
 = \tau^{-n}(\phi_1\!\times \!n
 +\phi_2)\times\left(1+\mathcal{O}\left(B^n\right)\right).
\end{equation}
It follows then immediately that
\begin{equation}
\label{eq:finHAC}
\Ex\left(\widetilde{H}_n^{(\mathtt{A}\rightarrow\mathtt{C})}\right)=\Ex(H_n^{(\mathtt{A}\rightarrow\mathtt{C})})\Big/\overline{f}_n=(c_1\times
n +
c_2)\times\left(1+\mathcal{O}\left(B^n\right)\right),\qquad (B<1).
\end{equation}
In the more general case, we have, for $n \ll
\min^{-1}_{\alpha,\beta\in \CA}\left(p_{\alpha\rightarrow \beta}\right)$,
\[
\mathfrak{p}_n\approx\sum_{\substack{\alpha\in\mathcal{A},\beta\in\mathcal{A}\\\alpha\neq\beta}}
\Ex\left(\widetilde{H}_n^{(\alpha\rightarrow\beta)}\right)\times
p_{\alpha\rightarrow \beta}(1)= (C_1\times n
+C_2)
\times\left(1+\mathcal{O}\left(K^n\right)\right),\quad (K<1),
\]
where $C_1$ and $C_2$ are constants, and $K$ is the maximum of the $|\mathcal{A}|(|\mathcal{A}|-1)$  constants $B$ used
when applying the Equation \eqref{eq:finHAC} to the  $|\mathcal{A}|(|\mathcal{A}|-1)$
types of mutation.

This proves Proposition~\ref{prop:linear}.
\ \vspace*{1cm}

\begin{figure}[!ht]
\setlength{\unitlength}{1.1mm}
\centering\begin{picture}(20,40)(0,0)
\put(-51,0)    {\includegraphics[height=4.8cm,width=4.8cm]{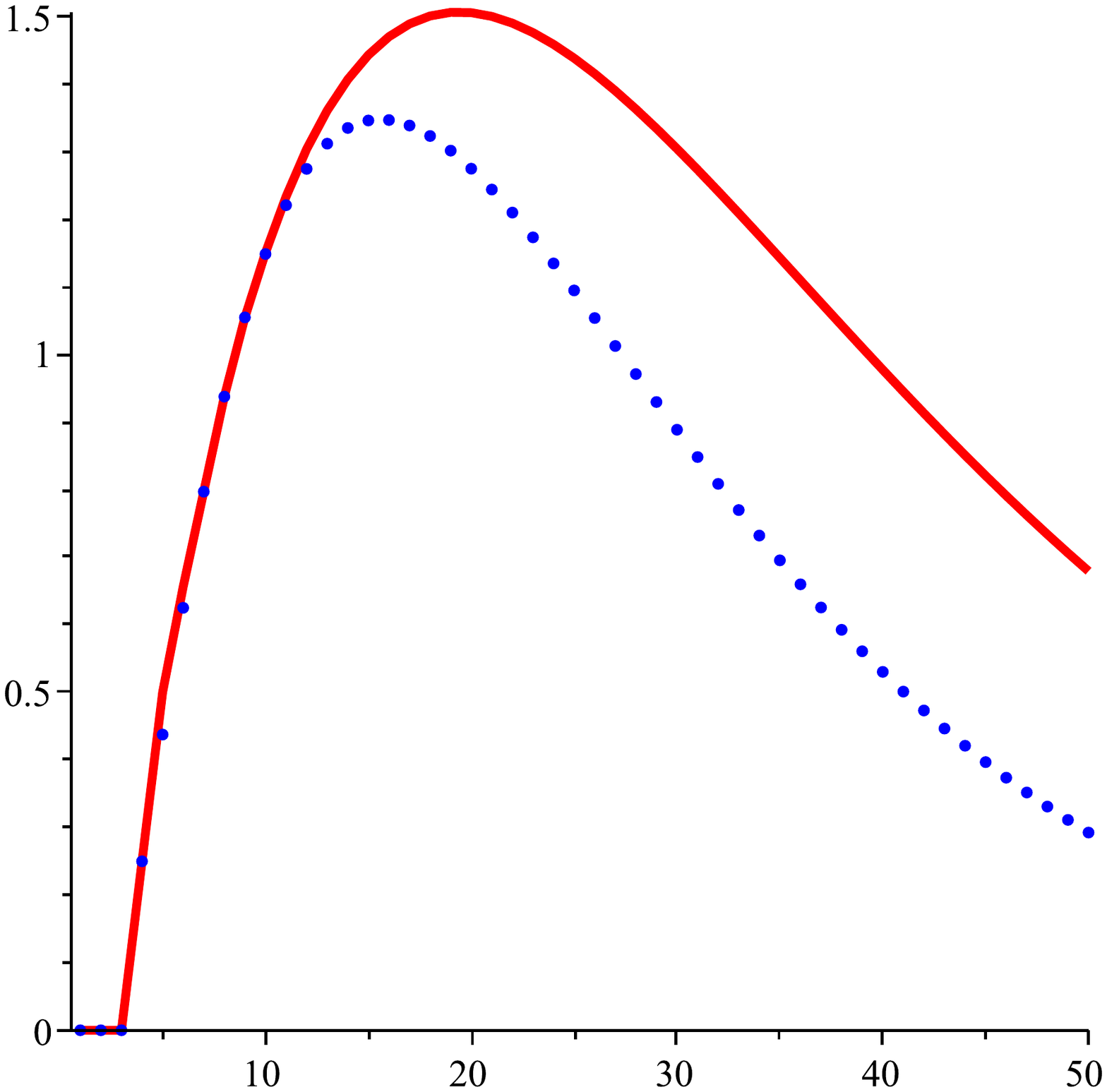}}
\put(-11.5,0)  {\includegraphics[height=4.8cm,width=4.8cm]{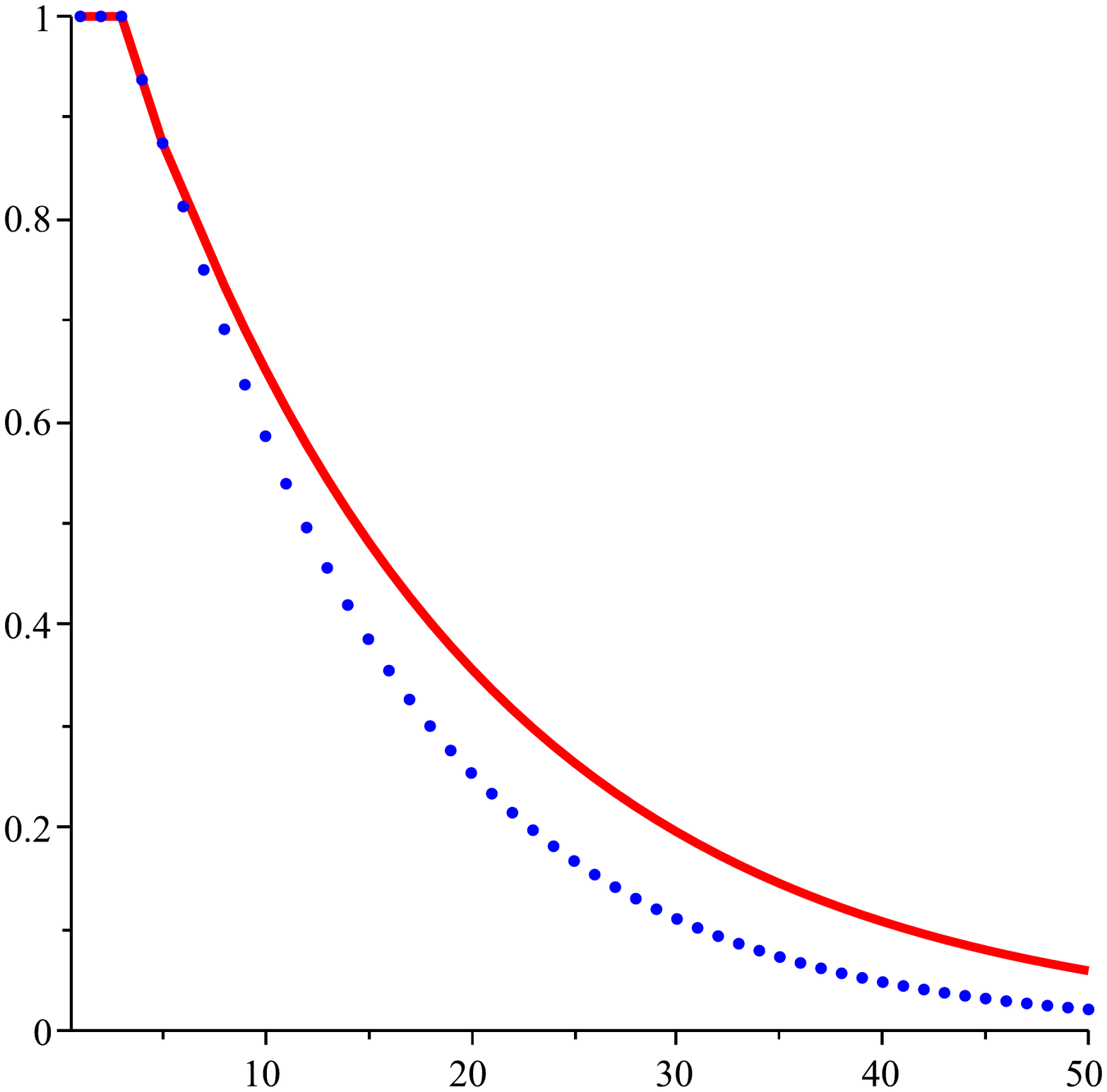}}
\put(28,0){\includegraphics[height=4.8cm,width=4.8cm]{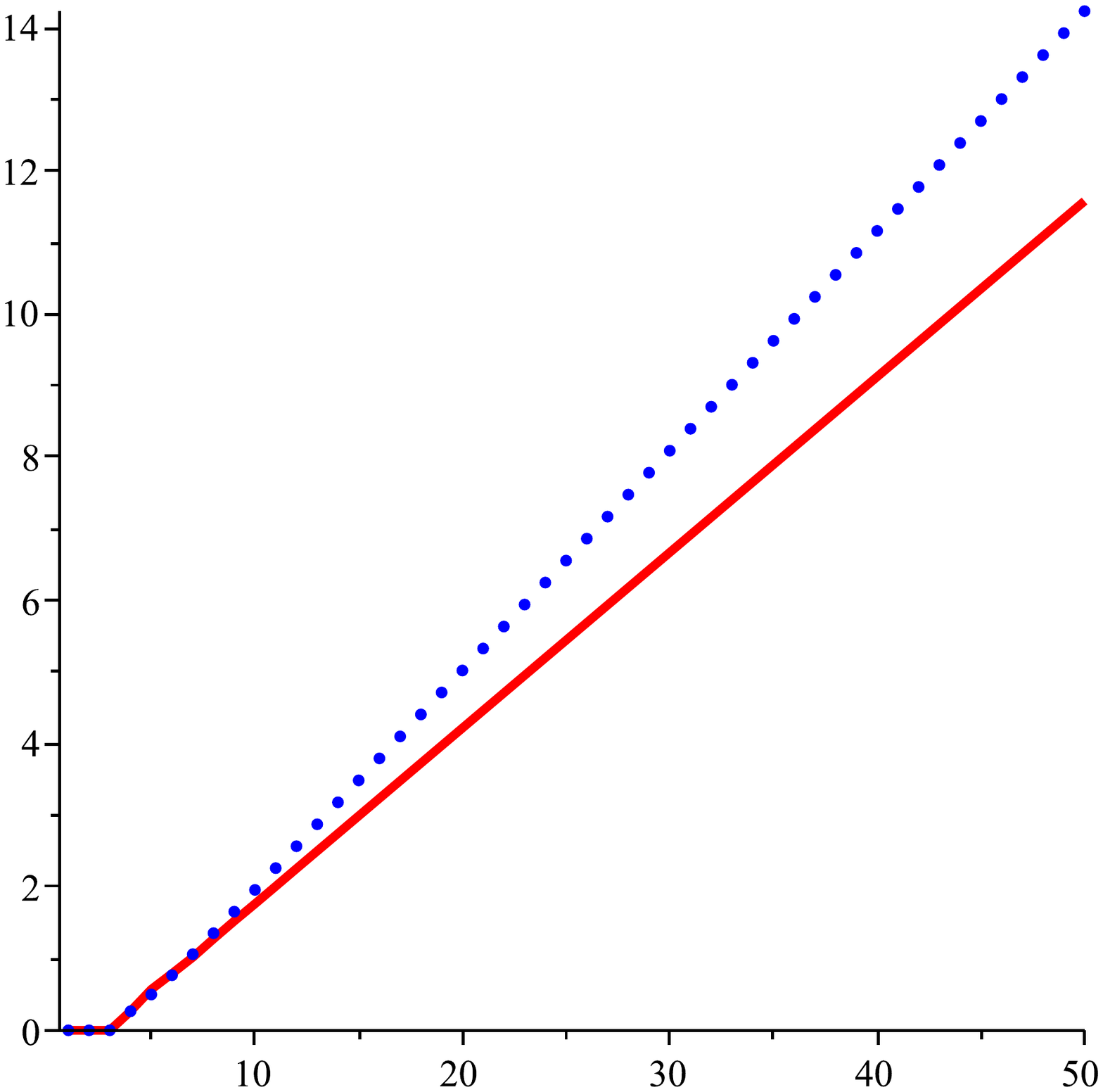}}
\put(-12,11){\scriptsize $n$}
\put(27,13){\scriptsize $n$}
\put(67,11){\scriptsize $n$}
\put(-48,39){\scriptsize $\eta_n=\Ex(H_n^{(\mathtt{A}\rightarrow\mathtt{C})})+\Ex(H_n^{(\mathtt{C}\rightarrow\mathtt{A})})$}
\put(-6,39){\scriptsize
  $\overline{f}^{(y)}_n=\Pr(|S_n(0)|_{y}=0)$}
\put(2,36){\tiny $(y \text{ is } b \text{ or }b')$}
\put(31,39){\scriptsize $\widetilde{\eta}_n=\Ex(\widetilde{H}_n^{(\mathtt{A}\rightarrow\mathtt{C})})+\Ex(\widetilde{H}_n^{(\mathtt{C}\rightarrow\mathtt{A})})$}
\end{picture}
\setlength{\unitlength}{1cm}
\begin{center}
\begin{picture}(0,0)
\put(-6.3,5.7){\rule{\textwidth}{0.2mm}}
\end{picture}
\end{center}
\ \vskip-9ex
\caption{\label{fig:GammaKM}{\small Asymptotic linear behaviour of the
    unconditionned $\eta_n$ (left) and conditionned
    $\widetilde{\eta}_n=\eta_n/\overline{f}_n^{(y)}$ (right)
 expectations of the number of
 putative-hit positions for
$b=\texttt{ACAC}$ (plain red lines) and $b'=\texttt{AACC}$ (blue
  dots) with the alphabet $\{\texttt{A},\texttt{C}\}$ and $\Pr(\texttt{A})=\Pr(\texttt{C})=1/2$.
See Equations \eqref{eq:fbarn}, \eqref{eq:Mn} and \eqref{eq:finHAC}}.
}
\end{figure}

\section{Toy example for the clump approach by language}
\label{sec:toyclump}
We  consider the following toy example\\[-2ex]
\[\mathcal{A}=\{\mathtt{A},\mathtt{C}\},\quad b=\texttt{AAA}, \quad
b'=\texttt{ACC},\qquad
 \Pr(\mathtt{A})=\Pr(\mathtt{C})=\frac{1}{2}.\]\ \\[-2ex]
We want to estimate the expectations of the {\sl total} number of putative-hit positions
$(\mathtt{A}\!\rightarrow\!\mathtt{C})$ and $(\mathtt{C}\!\rightarrow\!\mathtt{A})$
for the words
$b$ and $b'$.

Equation
\eqref{eq:gfputpos} becomes, with $\overline{\mathcal{A}^n_b}$ the subset of sequences of
size $n$ of $\overline{\mathcal{A}^{\star}_b}$,\ \\[-1ex]
\begin{equation}
F_b(z,t)=\sum_{n\geq 0}\sum_{w\in\overline{\mathcal{A}^n_b}}\frac{z^{n}}{2^n}t^{\operatorname{put-hit-pos}(w)}
\end{equation}
As mentioned earlier,  putative-hit positions only occur in the clumps, and
therefore the \rule{0pt}{8pt}core of differences between the behaviour of the
$3$-mers
$b=\texttt{AAA}$ and $b'=\texttt{ACC}$ come from differences in \rule{0pt}{10pt}the matrices of codes
$\bK_b$ and 
$\bK_{b'}$.\ \\[0.3ex]
\ \hspace*{0.4cm} We have
\scriptsize
\begin{center}
\begin{tabular}{l|l}
$b=\texttt{ACAC}$ & $b'=\texttt{AACC}$\\
$ d_{\ell}(b)=(\texttt{AAAC},\texttt{ACAA},\texttt{ACCC},\texttt{CCAC})$ & $ d_{\ell}(b')=(\texttt{AAAC},\texttt{AACA},\texttt{ACCC},\texttt{CACC})$ \\[2ex]
${\displaystyle \mathbb{K}_b=
   \left(\begin{array}{cccc}
          0 & \frac{z^2t}{4} &\frac{z^2t}{4} &\frac{z^3t}{8} \\[1ex]
          \frac{z^3t}{8}\!+\!\frac{z^2}{2}& \frac{z^3t}{8}& \frac{z^3t}{8}& 0\\[1ex]
          0 & 0 & 0 & \frac{z^3t}{8}\!+\!\frac{z^2}{2}\\[1ex]
          0 & \frac{z^2t}{4}& \frac{z^2t}{4} & \frac{z^3t}{8}
   \end{array} \right)}$
 & 
${\displaystyle  \mathbb{K}_{b'}=
  \left(\begin{array}{cccc}
          0 & \frac{zt}{2} & 0 & 0 \\[1ex]
          \frac{z^3t}{8}& \frac{z^3t}{8}&0& \frac{z^2t}{4}\\[1ex]
          0 & 0 & 0 &  \frac{z^3t}{8}\\[1ex]
          0 & 0 & \frac{zt}{2} & \frac{z^3t}{8}
 \end{array} \right)}$\\
\multicolumn{2}{c}{$$}\\[2ex]
\multicolumn{2}{l}{$d_{\ell}(b)(z,t)=d_{\ell}(b')(z,t)=(\frac{z^4t}{16},\frac{z^4t}{16},\frac{z^4t}{16},\frac{z^4t}{16})$}
\end{tabular}
\end{center}
\normalsize
\bigskip

\noindent
The last equations~\footnote{The extension
  $\texttt{AC}\in\CK_{\texttt{ACAA},\texttt{AAAC}}$ as in
  $\texttt{ACA}{\color{red}\underline{\texttt{A}}}|\texttt{AC}$
leads to {\sl no new}
  putative-hit position. The same remark applies to the extension
 $\texttt{AC}\in\CK_{\texttt{ACCC},\texttt{CCAC}}$.} intuitively suggests that
  there
should be  more putative-hit positions in a random
sequence
for $b=\texttt{ACAC}$ than for $b'=\texttt{AACC}$.
This is verified in Figure \ref{fig:GammaKM} (left) where we plot the
unconditionned expectations  of the number of putative-hit positions  for both
$4$-mers.
However, when conditionning 
as in  Figure \ref{fig:GammaKM} (right),
the $4$-mer $\texttt{ACAC}$ get lower expectations than 
the $4$-mer $\texttt{AACC}$; this follows from the  values of the constants
$C_1$ for $b$ and $b'$ that  respectively are $C_1=0.2452503893$ for
$b=\texttt{ACAC}$ and $C_1=0.3068491678$ for $b'=\texttt{AACC}$.

Figure~\ref{fig:GammaKM} moreover exhibits
 the linear behaviour 
of these expectations  with respect to the length $n$ of the sequences,
as stated in Proposition~\ref{prop:linear}.

\end{document}